\documentclass[aps, prl, reprint, letterpaper, longbibliography, superscriptaddress]{revtex4-1}

\pdfoutput=1

\usepackage{amsmath,amssymb,amsfonts,latexsym,color,dcolumn,bm,mathtools,amsbsy}
\usepackage[english]{babel}
\usepackage{graphicx}
\usepackage[utf8]{inputenc}
\usepackage{textcomp}
\usepackage[babel=true,final=true,protrusion=true, expansion=true]{microtype}
\usepackage[colorlinks,urlcolor=blue,citecolor=blue,unicode]{hyperref}
\hypersetup{
	pdftitle={Spatially Adiabatic Frequency Conversion in Optoelectromechanical Arrays},
	pdfauthor={O. \v{C}ernot\'ik, S. Mahmoodian, K. Hammerer}
}

\newcommand{\dd}{\ensuremath{d}}
\newcommand{\dt}{\ensuremath{\dd{}t}}
\newcommand{\dz}{\ensuremath{\dd{}z}}
\newcommand{\nbar}{\ensuremath{\bar{n}}}
\newcommand{\Hint}{\ensuremath{H_\mathrm{int}}}
\newcommand{\Hc}{\ensuremath{\mathrm{H.c.}}}
\newcommand{\diag}{\ensuremath{\mathrm{diag}}}
\newcommand{\vect}[1]{\ensuremath{\mathbf{#1}}}
\newcommand{\gvec}[1]{\ensuremath{\boldsymbol{#1}}}
\newcommand{\const}{\ensuremath{\mathrm{const.}}}

\begin{document}

\title{Spatially Adiabatic Frequency Conversion in Optoelectromechanical Arrays}

\author{Ond\v{r}ej \v{C}ernot\'ik}
\email{ondrej.cernotik@mpl.mpg.de}
\affiliation{Institute for Theoretical Physics, Institute for Gravitational Physics (Albert Einstein Institute), Leibniz University Hannover, Appelstra\ss{}e 2, 30167 Hannover, Germany}
\affiliation{Max Planck Institute for the Science of Light, Staudtstra\ss{}e 2, 91058 Erlangen, Germany}

\author{Sahand Mahmoodian}
\affiliation{Institute for Theoretical Physics, Institute for Gravitational Physics (Albert Einstein Institute), Leibniz University Hannover, Appelstra\ss{}e 2, 30167 Hannover, Germany}

\author{Klemens Hammerer}
\affiliation{Institute for Theoretical Physics, Institute for Gravitational Physics (Albert Einstein Institute), Leibniz University Hannover, Appelstra\ss{}e 2, 30167 Hannover, Germany}

\begin{abstract}
    Faithful conversion of quantum signals between microwave and optical frequency domains is crucial for building quantum networks based on superconducting circuits.
    Optoelectromechanical systems, in which microwave and optical cavity modes are coupled to a common mechanical oscillator, are a promising route towards this goal.
    In these systems, efficient, low-noise conversion is possible using mechanically dark mode of the fields,
    but the conversion bandwidth is limited to a fraction of the cavity linewidth.
    Here, we show that an array of optoelectromechanical transducers can overcome this limitation and reach a bandwidth that is larger than the cavity linewidth.
    The coupling rates are varied in space throughout the array so that the mechanically dark mode of the propagating fields adiabatically changes from microwave to optical or vice versa.
    This strategy also leads to significantly reduced thermal noise with the collective optomechanical cooperativity being the relevant figure of merit.
    Finally, we demonstrate that the bandwidth enhancement is, surprisingly, largest for small arrays; this feature makes our scheme particularly attractive for state of the art experimental setups.
\end{abstract}

\date{\today}

\maketitle

\emph{Introduction.}---Superconducting circuits are among the best platforms for quantum computing~\cite{Blais2007,Devoret2013}.
Strong nonlinearities in these systems are provided by Josephson tunnel junctions, precise control is possible using microwave signals, and advanced fabrication methods enable scaling their size up.
Experiments in recent years demonstrated quantum gates with several qubits \cite{Majer2007,Sillanpaa2007,Mariantoni2011,Barends2014}, basic quantum algorithms \cite{DiCarlo2009,DiCarlo2010,Fedorov2011}, and quantum error correction \cite{Reed2012,Corcoles2015,Kelly2015,Riste2015}.
Further scaling will require connecting superconducting circuits into quantum networks \cite{Kimble2008};
although short-distance communication is possible at microwave frequencies \cite{Vermersch2017,Xiang2017}, 
large networks will require interfacing superconducting systems with light.

\begin{figure}[b]
    \centering
    \includegraphics[width=\linewidth]{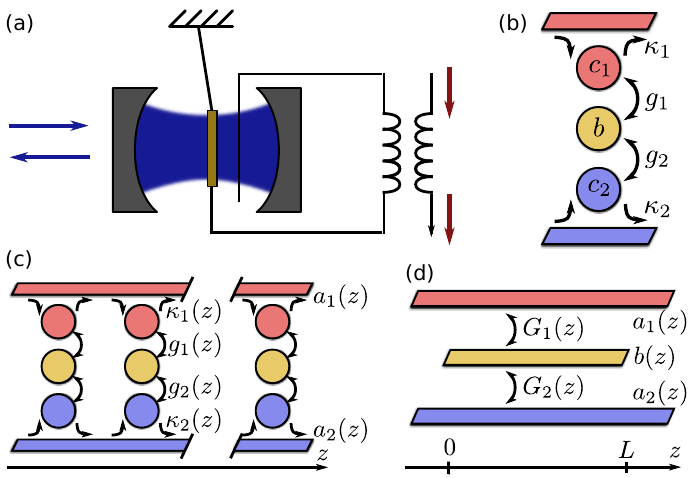} 
    \caption{\label{fig:Transducer}(Color online)
        (a) Basic optoelectromechanical transducer formed by coupling an optical cavity and a microwave resonator to a common mechanical oscillator
        (the yellow membrane in the middle of the optical cavity).
        (b) Schematic representation of the transducer, including waveguides for input and output fields.
        (c) Transducer array for spatially adiabatic frequency conversion.
        The transducers are directionally coupled; signals propagate from left to right.
        (d) Continuous model for frequency conversion where the propagating fields are coupled via a spatially extended mechanical mode $b(z)$.
    }
\end{figure}

As a result, transduction of quantum signals has attracted attention as an important task for quantum technologies \cite{Zeuthen2016} and
various systems have been proposed as suitable candidates for mediating interaction between microwaves and light:
Atomic, molecular, and solid-state impurity spins \cite{Sorensen2004,Tian2004,Rabl2006,OBrien2014,Xia2015,Das2017,Gard2017,Lekavicius2017}, magnons in ferromagnetic materials \cite{Hisatomi2016},
electrooptic modulators \cite{Tsang2010,Tsang2011,Javerzac-Galy2015,Rueda2016},
and mechanical oscillators \cite{Stannigel2010,Taylor2011,Barzanjeh2012,Tian2012,Wang2012,Clader2014,Yin2015,Cernotik2016,Okada2017} are all capable of interacting with both frequency domains.
Particularly optoelectromechanical systems [see Fig.~\ref{fig:Transducer}(a,b)] emerged as a promising and versatile platform with several experiments demonstrating efficient conversion between microwave and optical signals \cite{Bochmann2013,Andrews2014,Bagci2014,Balram2016}.

Optomechanical interaction is provided by radiation pressure while electromechanical coupling is due to electrostatic forces \cite{Aspelmeyer2013};
by coupling an optical cavity and a microwave resonator to the same mechanical oscillator, we can build a transducer for frequency conversion between these two frequency domains.
Various strategies have been proposed to fulfil this task with two---based on mechanically dark mode of the electromagnetic fields---especially resilient against mechanical noise \cite{Tian2012,Wang2012,Dong2012,Hill2012,Andrews2014}:
In the first approach, time-independent interaction is used to convert propagating fields using an effect akin to optomechanically induced transparency \cite{Weis2010}.
This setup is easy to implement and is capable of converting arbitrary input signals but reaches only a limited conversion bandwidth (given by the optically broadened mechanical linewidth, typically much smaller than the cavity linewidth).
The spectral width can be increased by using the second strategy---converting intracavity fields by adiabatic passage.
This scenario, however, works only with a single temporal mode, requires time-dependent control, pulse shaping of the incoming signals to store them in the cavity, and strong optomechanical coupling; all these demands make its experimental implementation much more challenging.
The process can be sped up using shortcuts to adiabaticity \cite{Zhou2016,Baksic2017} (thus relaxing requirements on the coupling strength),
but these techniques require correction Hamiltonians to compensate for non-adiabatic transitions and lead to more complex time control schemes.
Finally, the bandwidth can be enhanced when the conversion is accompanied by cross-amplification~\cite{Ockeloen-Korppi2016};
in this case, however, the signal also gets amplified in the process which, depending on the application, might be an undesired side effect.

Here, we prove that the limitation on conversion bandwidth can be overcome in an array of optoelectromechanical transducers and that frequency conversion of multimode signals over a bandwidth larger than the cavity linewidth is possible.
In the system, depicted in Fig.~\ref{fig:Transducer}(c), the nature of the mechanically dark mode of the propagating fields is varied in space rather than in time;
the propagating signal is adiabatically converted from one propagating field (e.g., microwave) to the other (optical).
We demonstrate that the bandwidth can be enhanced by this strategy and can surpass the cavity linewidth;
simultaneously, added noise---coming from the thermal mechanical reservoir---is strongly suppressed.
Strategies based on spatial adiabatic passage have already been used for frequency conversion, mode splitting, and spectral filtering in the optical domain~\cite{Suchowski2008,Tseng2010,Menchon-Enrich2013}.
In contrast, we use adiabatic dynamics with spatially varying parameters to bridge two vastly different frequency domains (microwaves and light);
additionally, we show that our approach brings advantage also for small transducer arrays where the adiabatic condition is not fulfilled.

\emph{Continuous model.}---The strategy for adiabatic conversion of propagating signals without time-dependent control is best explained by considering a spatially extended structure, in which the coupling rates can be varied in space, rather than in time; cf. Fig.~\ref{fig:Transducer}(d).
We can describe such an interaction with the following model:
Two 1D fields of propagating photons [with annihilation operators $a_{1,2}(z)$] couple over a length $L$ to a 1D field of phonons [annihilation operator $b(z)$] via beam splitter interactions at strengths $G_{1,2}(z)$. The Hamiltonian of the full system is thus $H = H_0 + \Hint$ with the free Hamiltonian and interaction \cite{Zoubi2016}
\begin{subequations} 
\begin{align}\label{eq:H0}
    H_0 &= -i\int_{-\infty}^\infty\dz\sum_{i=1}^2 v_ia_i^\dagger(z)\frac{\partial}{\partial z} a_i(z) + \Hc,\\
    \label{eq:coupling}
    \Hint &= \int_0^L\dz[G_1(z)a_1^\dagger(z) + G_2(z)a_2^\dagger(z)]b(z) + \Hc \nonumber\\
    & = \int_0^L\dz G[d_1^\dagger(z)b(z) + b^\dagger(z)d_1(z)];
\end{align}
\end{subequations}
$H_0$ describes propagation of photons of type $i=1,2$ in the positive $z$-direction at speed $v_i$. In the interaction, we introduced the propagating normal mode $d_1(z) = [G_1(z)a_1(z) + G_2(z)a_2(z)]/G(z)$ with $G^2(z) = G_1^2(z)+G_2^2(z)$.
The orthogonal normal mode $d_2(z) = [G_2(z)a_1(z) - G_1(z)a_2(z)]/G(z)$ is not directly coupled to the mechanical mode.
To ensure that the mode $d_2$ remains mechanically dark, we have to confirm that the normal modes are not coupled in the free Hamiltonian $H_0$.

To derive conditions under which the normal modes stay decoupled, we collect the propagating fields in a vector $\vect{a}(z) = [a_1(z),a_2(z)]^T$;
the normal modes are collected in a similar vector $\vect{d}(z)$.
The transformation between the propagating fields and normal modes can be described by the orthogonal matrix $\vect{O}(z)$,
\begin{equation} 
    \vect{a}(z) = \vect{O}(z)\vect{d}(z) = G^{-1}\left(\begin{array}{cc} G_1&G_2\\G_2&-G_1
    \end{array}\right)
    \left(\begin{array}{c} d_1\\d_2 \end{array}\right).
\end{equation}
Plugging this expression into the free Hamiltonian \eqref{eq:H0} [which we write in the matrix form $H_0 = -i\int\dz\vect{a}^\dagger\vect{V}\partial_z\vect{a}+\Hc$ with $\vect{V} = \diag(v_1,v_2)$],
we find that the normal modes remain decoupled if the matrices $\vect{O}^T\vect{VO}$ and $\partial_z\vect{O}^T(z)\vect{VO}(z)$ are diagonal.
The matrix $\vect{O}^T\vect{VO}$ is diagonal if both fields propagate at the same velocity, $v_1=v_2 = v$;
the other matrix is, under this condition, identically zero.
The normal modes thus remain decoupled and the mode $d_2$ is a dark mode of the dynamics.
Although this result has been, for clarity, derived for linear dispersion, one can easily generalize it to arbitrary dispersion of the travelling fields;
the dark mode exists as long as both fields have the same dispersion.

For concreteness, we consider conversion from the microwave field $a_1$ to the optical field $a_2$ which is achieved by varying the coupling strengths from $G_1(0)\ll G_2(0)$ to $G_1(L)\gg G_2(L)$ such that an incoming microwave photon $a_1(0)=d_2(0)$ is converted adiabatically into an optical photon $d_2(L)=a_2(L)$. 
To ensure that the conversion stays adiabatic, the change of the coupling rates has to be slow so that we do not excite the orthogonal normal mode $d_1$.
For the usual temporal adiabatic passage, the condition $|\dd{}G_i/\dt|\ll G^2$ can be derived \cite{Tian2012}.
Upon rewriting the time derivative as a spatial derivative, the adiabatic condition becomes $|{\dd{}G_i}/{\dz}|\ll {G^2}/{v}$.

\emph{Transducer array.}---The continuous dynamics can be approximated in an array of optoelectromechanical transducers; cf. Fig.~\ref{fig:Transducer}(c).
Each transducer is formed by a mechanical oscillator coupled to an optical and a microwave cavity;
the interaction is described by the interaction Hamiltonian $\Hint = g_1(c_1^\dagger b+b^\dagger c_1) + g_2(c_2^\dagger b+b^\dagger c_2)$ (with cavity modes $c_{1,2}$ and coupling rates $g_{1,2}$).
This form of interaction can be obtained from the standard linearized optomechanical coupling when the cavity modes are driven on the lower mechanical sideband \cite{Aspelmeyer2013};
in the resolved sideband regime, $\kappa_i\ll\omega_m$ (i.e., when the cavity linewidths $\kappa_i$ are smaller than the mechanical frequency $\omega_m$), we can apply the rotating wave approximation and obtain the beam splitter Hamiltonian crucial for state transfer.
The dynamics is governed by the Heisenberg-Langevin equations
\begin{subequations}\label{eq:HLEq} 
\begin{align}
    \dot{c}_i &= -\frac{\kappa_i}{2}c_i - ig_ib + \sqrt{\kappa_i}a_i(z_j^-), \\
    \dot{b} &= -\frac{\gamma}{2}b - ig_1c_1-ig_2c_2 + \sqrt{\gamma}b_\mathrm{in}
\end{align}
\end{subequations}
with the input-output relations $a_i(z_j^+) = \sqrt{\kappa_i}c_i - a_i(z_j^-)$;
we denote the input and output fields of the transducer element at position $z_j$ by $a_i(z_j^-)$ and $a_i(z_j^+)$, respectively. The thermal Langevin force acting on the mechanical oscillator is $b_\mathrm{in}$ and the mechanical linewidth is $\gamma$.

\begin{figure}[t]
    \centering
    \includegraphics[width=\linewidth]{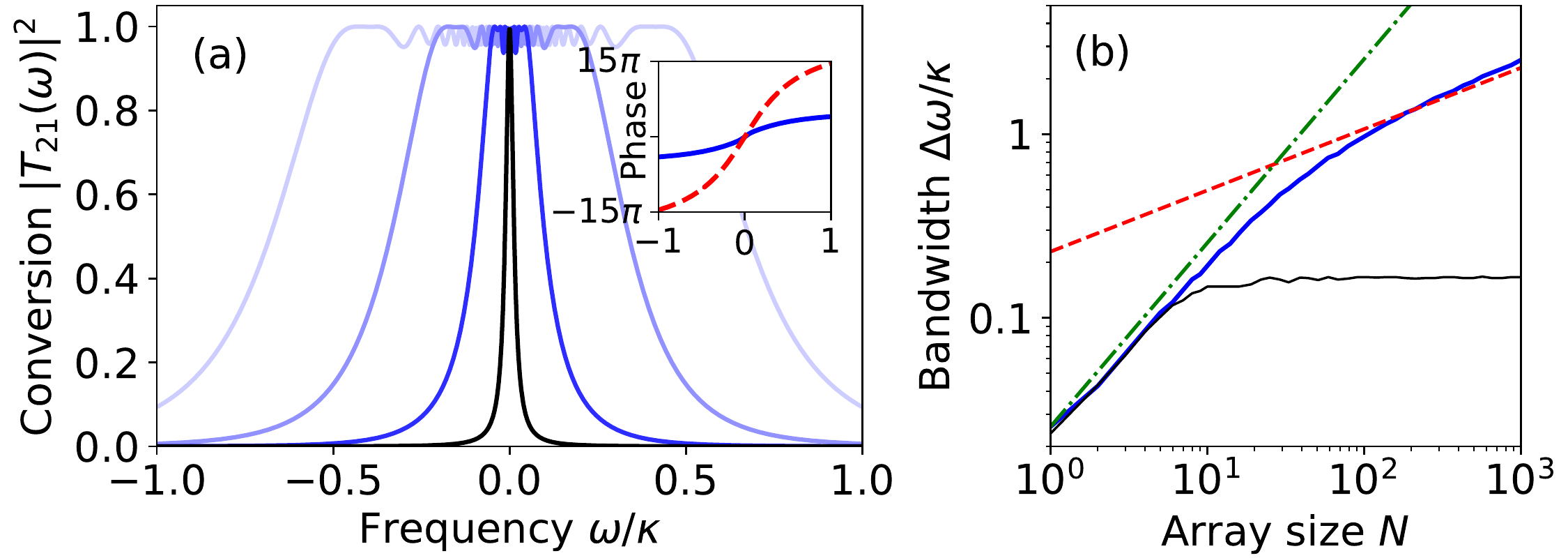} 
    \caption{\label{fig:TransferMatrix}(Color online)
        (a) Energy conversion spectrum for different array sizes.
        From dark colors to light, the array size is $N = 1, 10, 50, 200$.
        The inset shows the phase of the conversion coefficient versus frequency for array sizes $N = 5$ (solid blue line) and $N = 20$ (dashed red line).
        (b) Transducer bandwidth (full-width half-maximum of the energy conversion coefficient $|T_{21}(\omega)|^2$) versus array size.
        The solid blue line shows results of numerical simulations;
        the dashed red line represents the analytical result~\eqref{eq:bandwidth}
        and the dot-dashed green line the linear fit $4g^2N/\kappa$.
        With the thin black line, we plot the bandwidth for an asymmetric transducer with $\kappa_1 = \kappa = \kappa_2/10$.
        For both panels (and following discussion), we use $g = 0.08\kappa$.
    }
\end{figure}

The Heisenberg-Langevin equations \eqref{eq:HLEq} can be solved in frequency domain, which enables us to describe the relation between the input and output fields by the scattering matrix, 
$\vect{a}(z_j^+,\omega) = \vect{S}_j(\omega)\vect{a}(z_j^-,\omega)$.
We obtain the transfer through the whole array by multiplying the scattering matrix of all transducers,
$\vect{a}(L,\omega) = \vect{S}_N(\omega)\vect{S}_{N-1}(\omega)\ldots\vect{S}_1(\omega)\vect{a}(0,\omega) = \vect{T}(\omega)\vect{a}(0,\omega)$.
Frequency conversion from microwaves to light is characterized by the matrix element $T_{21}(\omega)$ of the resulting scattering matrix $\vect{T}(\omega)$; see Fig.~\ref{fig:TransferMatrix}(a) for an illustration.
In this description, we drop the effect of thermal noise coming from the mechanical reservoir;
we discuss its role further below.

Conversion via the mechanically dark mode is achieved by varying the coupling rates from $g_1(0)\approx 0$, $g_2(0)=\bar{g}_2$ at the beginning of the array to $g_1(L)=\bar{g}_1$, $g_2(L)\approx 0$ at its end;
this ensures that the mode varies from $d_2(0) = a_1$ to $d_2(L) = a_2$.
The condition of equal propagation velocities implies that the two fields have to acquire the same phase in propagation between two sites.
(We neglected free propagation in the description above but it can be included in the transfer matrix formalism; see the Supplemental Material \cite{Supplement}.)
Similarly, the adiabatic condition is fulfilled for $\bar{g}_i\sqrt{N} > \kappa_i$;
this result follows from eliminating the cavity fields, from which we obtain $G_i = g_i^2/\kappa_i$.

Frequency conversion with this strategy is efficient only for a limited range of frequencies.
We can find the bandwidth $\Delta\omega$ (i.e., the frequency width of the conversion coefficient $|T_{21}(\omega)|^2$) from the following consideration:
Far off resonance, the probability of a transducer converting a photon is small and proportional to $g_1g_2\kappa/\omega^3\ll 1$.
In an array, the probability is enhanced by sending the signal through $N$ transducers;
the conversion efficiency scales as $\bar{g}_1\bar{g}_2\kappa N/\omega^3$.
We can therefore expect the conversion bandwidth to grow with the cubic root of the array size.
In the Supplemental Material \cite{Supplement}, we derive the bandwidth rigorously from the transfer matrix and show that, for a symmetric array ($\bar{g}_1 = \bar{g}_2 = g$, $\kappa_1 = \kappa_2 = \kappa$),
\begin{equation}\label{eq:bandwidth} 
    \Delta\omega = \left(\frac{4\sqrt{2}}{3} g^2\kappa N\right)^{1/3}.
\end{equation}
This finding presents the first main result of our paper:
In the adiabatic limit, $g\sqrt{N}>\kappa$, the conversion bandwidth becomes larger than the cavity linewidth, $\Delta\omega>\kappa$,
which is a dramatic enhancement compared to a single transducer for which $\Delta\omega \propto g^2/\kappa\ll\kappa$ \cite{Hill2012}.

\emph{Numerical simulations.}---Results of the transfer matrix analysis are shown in Fig.~\ref{fig:TransferMatrix}.
Panel~(a) shows the energy conversion coefficient $|T_{21}(\omega)|^2$ as a function of frequency for an increasing array size (from dark to light).
As the array size increases, the dynamics better approximates the continuous adiabatic state transfer, resulting in larger conversion bandwidth.
This result is further accentuated in panel~(b), where the bandwidth is shown versus array size.
In the large-array limit, the bandwidth indeed agrees with the analytical formula given by Eq.~\eqref{eq:bandwidth}.
During conversion, the signal acquires a large phase shift [shown in the inset of Fig.~\ref{fig:TransferMatrix}(a)] owing to reflection from a large number of cavities;
the phase across the whole frequency spectrum grows linearly with array size and is equal to $2\pi N$.
In practical applications, this phase shift has to be taken into account in postprocessing or compensated by a suitable phase shift on the input or output field.

Interestingly, the conversion bandwidth is enhanced also when the adiabatic condition is not fulfilled.
In this case, the scaling of bandwidth is even more favorable---close to linear in the array size, as shown in Fig.~\ref{fig:TransferMatrix}(b).
This result can be understood by noting that for frequencies within the cavity linewidth, $\omega<\kappa$, the cavity fields can be adiabatically eliminated from the dynamics;
the probability of converting a single photon in a transducer is then inversely proportional to the frequency, $p_1\propto\omega^{-1}$.
This is our second main result: The largest enhancement of the conversion bandwidth (per transducer element) is achieved in small arrays, making our strategy particularly promising for near-future experimental implementations. 
As the bandwidth further increases and approaches the cavity linewidth (so that the fields cannot be adiabatically eliminated), the scaling changes from linear to $N^{1/3}$ dependence.

When the optical and microwave decay rates differ, $\kappa_1\neq\kappa_2$, Eq.~\eqref{eq:bandwidth} cannot be applied; the conversion bandwidth saturates [cf. Fig.~\ref{fig:TransferMatrix}(b)].
This behavior is a consequence of phase mismatch between the two propagating fields:
The different decay rates result in different dispersion of the propagating fields such that a mechanically dark mode no longer exists.
The use of a transducer array, however, still provides an improved bandwidth compared to a single optomechanical transducer.
In addition, we discuss in the Supplemental material~\cite{Supplement} a simple strategy based on spatial variation of the decay rates which can recover the cubic-root scaling when both decay rates are of the same order.

\emph{Losses and noise.}---To fully characterize frequency conversion, we need to determine not only the conversion efficiency and bandwidth but also quantify the noise added in the process \cite{Zeuthen2016}.
To limit thermal noise in the microwave field, the whole device (i.e., the whole transducer array) should be placed in a single cryostat; the microwaves are then effectively at zero temperature.
Any residual thermal occupation will slightly elevate the noise floor as is typical for microwave experiments at cryogenic temperatures.
The main source of noise is then the thermal bath of the mechanical oscillators.
For a single transducer, the spectral density of the added noise scales as $1/C = (4g^2/\kappa\gamma\nbar)^{-1}$ (with $\nbar$ being the thermal occupation of the bath) \cite{Wang2012};
in an array, we can expect the noise to be enhanced by the array size $N$.
On the other hand, our conversion strategy uses a mechanically dark mode of the propagating fields and is thus protected against mechanical noise.
The noise amplitude is suppressed by the adiabaticity parameter $1/N$ [we obtain this result from the expression $(dg/dz)/g$ with linear variation of the coupling rates].
The noise spectral density is then suppressed by the square of the adiabatic parameter and the total added noise is proportional to $1/CN$;
we prove this statement rigorously in the Supplemental Material \cite{Supplement}.

This is our third main result:
In an optoelectromechanical array, the added noise is suppressed by the collective optomechanical cooperativity, $CN>1$, representing a large improvement compared to a single transducer where $C>1$ is needed.
Outside the adiabatic regime (i.e., in small arrays), thermal noise is still suppressed compared to a single transducer;
we discuss this effect in detail in the Supplemental Material~\cite{Supplement}.

The second source of noise is the Stokes scattering associated with the opto- and electromechanical interaction.
The full linearized interaction between a cavity field and a mechanical oscillator under a strong drive is described by the Hamiltonian $\Hint = g_i(c_i+c_i^\dagger)(b+b^\dagger)$.
If the cavity is driven on the lower mechanical sideband, we can apply the rotating wave approximation and obtain the beam splitter interaction necessary for state transfer.
This approximation (which neglects the heating associated with the two-mode squeezing part of the interaction) is valid if the device operates in the resolved-sideband regime, $\kappa_i\ll\omega_m$.
Furthermore, the approximation implies that Fourier frequencies of interest follow the same rule, $\omega\ll\omega_m$;
the mechanical frequency thus provides a limit on conversion bandwidth in practical realizations.
A more detailed analysis, showing that the Stokes scattering noise does not become prohibitive, is presented in the Supplemental Material~\cite{Supplement}.

\begin{figure}
    \centering
    \includegraphics[width=\linewidth]{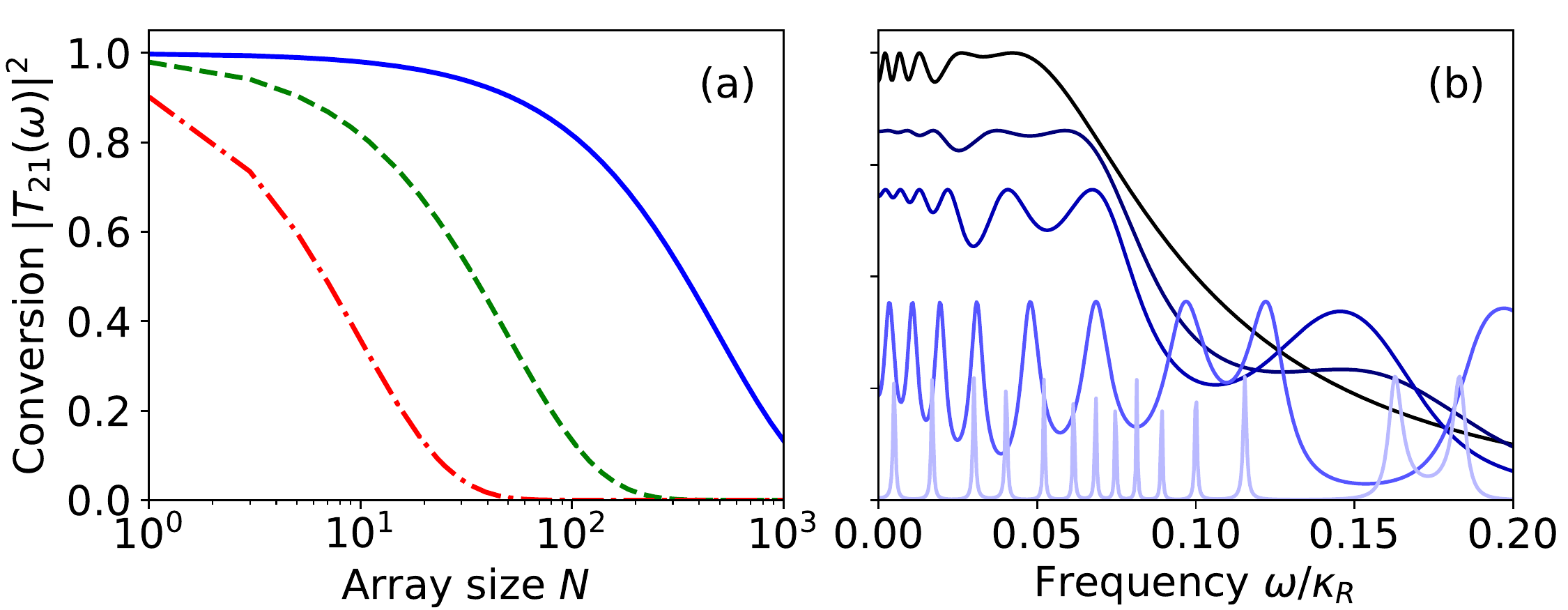} 
    \caption{\label{fig:Losses}(Color online)
    (a) Conversion efficiency (on resonance) versus array size for transmission loss between neighboring transducers 0.1 \% (solid blue line), 1 \% (dashed green line), and 5 \% (dot-dashed red line).
    (b) Conversion spectrum (its positive-frequency part) for $N = 10$ transducers with backscattering rates (from dark to light) $\kappa_L/\kappa_R = 0,\ 0.1, 0.2,\ 0.5,\ 0.9$.
    }
\end{figure}

Finally, we must also consider optical and microwave losses.
Electromagnetic fields can effectively decay via two distinct processes: by direct loss---in propagation or in cavities---and backscattering at cavity mirrors.
Direct losses are analogous to cavity loss in adiabatic conversion of intracavity fields;
efficient conversion requires them to be smaller than coupling of the propagating fields to the mechanical oscillators; cf. Fig.~\ref{fig:Losses}(a) for scaling of the conversion efficiency with transmission loss.
Intracavity losses are qualitatively similar;
we defer a detailed quantitative comparison of these two types of loss to the Supplemental Material~\cite{Supplement}.

Backscattering has no analog in the usual temporal adiabatic dynamics where it would correspond to signals propagating backwards in time.
To model this process, we assume that each cavity can decay into a right- or left-propagating field at a rate $\kappa_{R,L}$.
(Previously, we had $\kappa_R = \kappa$ and $\kappa_L = 0$.)
The backscattered (i.e., left-propagating) signal interferes with incoming signal, leading to reduction of conversion efficiency and interference pattern in the conversion spectrum.
This effect, which is captured in Fig.~\ref{fig:Losses}(b) and discussed in detail in the Supplemental Material \cite{Supplement}, 
reduces, to leading order, the conversion efficiency linearly with the backscattering rate, $\eta \approx 1-\alpha \kappa_L/\kappa_R$; from numerical simulations, we infer $\alpha\approx 1.6$.

\emph{Discussion and conclusions.}---Optoelectromechanical arrays for frequency conversion can be implemented in integrated systems with optomechanical crystals or microdisk optical resonators.
Both opto- and electromechanical interactions have been demonstrated with these systems \cite{Pitanti2015,Fink2016};
additionally, optomechanical arrays (albeit with photon hopping between sites and not directional propagation) have been constructed with whispering-gallery resonators \cite{Zhang2015}.
Variation of the coupling rates can be achieved by varying the single-photon coupling rates across the array---a single driving field can then be used for each type of cavities (i.e., one drive for microwaves and one for light; the pump fields will be subject to the same losses as the signal and, as discussed above, the losses have to be kept small for efficient conversion).
Building a large array would be extremely challenging, but even a small array is sufficient to considerably improve the conversion bandwidth.
With the relaxed conditions on optomechanical cooperativity, small arrays of integrated optomechanical transducers might soon be experimentally realizable and offer an $N$-fold enhancement of the conversion bandwdith compared to a single transducer.
Going beyond implementations in transducer arrays, it would be interesting to investigate whether direct conversion of propagating fields is possible in continuum systems \cite{Rakich2016,Zoubi2016}.

In summary, we demonstrated that the bandwidth of microwave-optical frequency conversion can be significantly enhanced in a one-dimensional optoelectromechanical array and is limited only by the mechanical frequency.
The strategy uses a mechanically dark mode of two propagating fields;
by varying the opto- and electromechanical coupling rates in space, we can achieve adiabatic conversion of signals between the two fields.
Our approach simultaneously leads to a siginifcantly reduced mechanical noise in the signal, even for weak optomechanical cooperativity;
since the coupling of the dark mode to the thermal mechanical environment is suppressed by the adiabaticity parameter, the relevant figure of merit for suppressing thermal noise is the collective optomechanical cooperativity.
Remarkably, efficient transduction with an improved bandwidth is also possible outside the adiabatic regime in small optoelectromechanical arrays;
the proposed strategy can thus be implemented with near-future quantum devices.


\begin{acknowledgments}
This work was funded by the European Commission (FP7-Programme) through iQUOEMS (Grant Agreement No. 323924).
We gratefully acknowledge support by DFG through QUEST and by the cluster system team at the Leibniz University Hannover.
\end{acknowledgments}

\bibliography{TransducerArray}

\clearpage
\setcounter{equation}{0}
\renewcommand \theequation {S\arabic{equation}}
\setcounter{figure}{0}
\renewcommand \thefigure {S\arabic{figure}}

\begin{widetext}

\begin{center}
    {\large\textbf{Supplemental Material: Spatially Adiabatic Frequency Conversion\\
        in Optoelectromechanical Arrays}}

    \vspace{0.2in}
    Ond\v{r}ej \v{C}ernot\'ik,$^{1,2}$ Sahand Mahmoodian,$^1$ and Klemens Hammerer$^1$

    \vspace{0.05in}
    $^1$\emph{Institute for Theoretical Physics, Institute for Gravitational Physics (Albert Einstein Institute),\\
    Leibniz University Hannover, Appelstra\ss{}e 2, 30167 Hannover, Germany}
    
    $^2$\emph{Max Planck Institute for the Science of Light, Staudtstra\ss{}e 2, 91058 Erlangen, Germany}

    (Dated: \today)
\end{center}

\setcounter{section}{0}
\renewcommand \thesection {S\roman{section}}

\section{Transfer matrix analysis}

We describe the propagation of signals through the transducer array using transfer matrix formalism.
The action of each transducer on the fields is described by its transfer (or scattering) matrix; we obtain the effect of the whole array by multiplying the transfer matrices of individual transducers.
For one-sided cavities, the scattering and transfer matrices are identical and both terms can be used interchangeably;
the distinction becomes important when describing the scattering of signals on two-sided cavities.

To find the transfer matrix of a single transducer, we solve the corresponding state-space model~\cite{Wang2012,Andrews2014}.
Two cavity modes (microwave and optical) interact with a mechanical oscillator via a beam splitter Hamiltonian
\begin{equation}
    H = g_1(c_1^\dagger b+b^\dagger c_1) + g_2(c_2^\dagger b+b^\dagger c_2).
\end{equation}
The dynamics of the transducer is characterized by the Heisenberg-Langevin equations, which we write in the matrix form
\begin{subequations}\label{eq:SLangevin}
\begin{align}
    \dot{\vect{a}} &= \vect{Aa} + \vect{Ba}_\mathrm{in}\label{eq:SLangevinHEq} \\
    \vect{a}_\mathrm{out} &= \vect{Ca} + \vect{Da}_\mathrm{in},
\end{align}
\end{subequations}
where $\mathbf{a} = (c_1,c_2,b)^T$, $\mathbf{a}_\mathrm{in} = [a_1(z^-),a_2(z^-)]^T$, and $\mathbf{a}_\mathrm{out} = [a_1(z^+),a_2(z^+)]^T$.
The matrices are given by
\begin{subequations}
\begin{align}
    \vect{A} &= \left(\begin{array}{ccc} -\displaystyle\frac{\kappa_1}{2}&0&-ig_1 \\ 0&-\displaystyle\frac{\kappa_2}{2}&-ig_2 \\ -ig_1&-ig_2&-\displaystyle\frac{\gamma}{2} \end{array}\right),\\
    \vect{B}^T = \vect{C} &= \left(\begin{array}{ccc} \sqrt{\kappa_1}&0&0 \\ 0&\sqrt{\kappa_2}&0 \end{array}\right),\\
    \vect{D} &= -\vect{I}_2;
\end{align}
\end{subequations}
here, $\vect{I}_2$ is the $2\times 2$ identity matrix.
In the frequency space, the relation between input and output fields is captured by the scattering matrix, $\vect{a}_\mathrm{out}(\omega) = \vect{S}(\omega)\vect{a}_\mathrm{in}(\omega)$, where
\begin{equation}
    \vect{S}(\omega) = \vect{D}-\vect{C}(\vect{A}+i\omega\vect{I}_3)^{-1}\vect{B},
\end{equation}
where the frequency $\omega$ is taken with respect to the cavity resonance.

The input and output fields contain only the propagating fields and not the mechanical bath.
This approach enables us to include losses in the propagating fields associated with the mechanical reservoir but does not include thermal mechanical noise that enters the fields (we will discuss this effect later).
The state-space model can be solved analytically; we obtain the scattering matrix
\begin{subequations}\label{eq:SMScattering}
\begin{align}
    \vect{S}(\omega) &= \left(\begin{array}{cc}\label{eq:SMScatterFull}
        -1 + \displaystyle\frac{8g_2^2\kappa_1 + 2\kappa_1(\kappa_2-2i\omega)(\gamma-2i\omega)}{D} & -\displaystyle\frac{8g_1g_2\sqrt{\kappa_1\kappa_2}}{D} \\ 
        -\displaystyle\frac{8g_1g_2\sqrt{\kappa_1\kappa_2}}{D} & -1 + \displaystyle\frac{8g_1^2\kappa_2 + 2\kappa_2(\kappa_1-2i\omega)(\gamma-2i\omega)}{D}       
    \end{array}\right) \\
    &\approx \left(\begin{array}{cc}
        \displaystyle\frac{\kappa_1+2i\omega}{\kappa_1-2i\omega} - \displaystyle\frac{8g_1^2\kappa_1}{(\kappa_1-2i\omega)^2(\gamma-2i\omega)} & -\displaystyle\frac{8g_1g_2\sqrt{\kappa_1\kappa_2}}{(\kappa_1-2i\omega)(\kappa_2-2i\omega)(\gamma-2i\omega)} \\ 
        -\displaystyle\frac{8g_1g_2\sqrt{\kappa_1\kappa_2}}{(\kappa_1-2i\omega)(\kappa_2-2i\omega)(\gamma-2i\omega)}  & \displaystyle\frac{\kappa_2+2i\omega}{\kappa_2-2i\omega} - \displaystyle\frac{8g_2^2\kappa_2}{(\kappa_2-2i\omega)^2(\gamma-2i\omega)}      
    \end{array}\right).
\end{align}
\end{subequations}
Here, $D = 4g_1^2(\kappa_2-2i\omega) + 4g_2^2(\kappa_1-2i\omega) + (\kappa_1-2i\omega)(\kappa_2-2i\omega)(\gamma-2i\omega)$ and
the second line approximates the scattering matrix in the weak-coupling regime, $g_i\ll\kappa_i$.
This approximation is not valid close to resonance, $\omega\approx 0$;
on resonance, we can express the scattering matrix using the classical cooperativities $\tilde C_i = 4g_i^2/\kappa_i\gamma$,
\begin{equation}\label{eq:SMScatterRes}
    \vect{S}(0) = \frac{1}{\tilde C_1+\tilde C_2+1}\left(\begin{array}{cc}
        -\tilde C_1+\tilde C_2+1 & -2\sqrt{\tilde C_1\tilde C_2} \\ -2\sqrt{\tilde C_1\tilde C_2} & \tilde C_1-\tilde C_2+1
    \end{array}\right).    
\end{equation}
We obtain the transfer matrix of the array by multiplying the scattering matrices of the transducers,
\begin{equation}
    \vect{T}(\omega) = \vect{S}_N(\omega)\vect{S}_{N-1}(\omega)\ldots\vect{S}_1(\omega);
\end{equation}
in this expression, $\vect{S}_j(\omega)$ is the scattering matrix of the $j$th transducer in the array.

\section{Conversion bandwidth in the adiabatic limit}

The transfer matrix formalism can be used to find the conversion bandwidth using the following approach:
We assume that the decay rates of the microwave and optical cavities are equal and constant across the whole array, $\kappa_1 = \kappa_2 = \kappa$.
We can then write the transfer matrix of the $j$th transducer as
\begin{equation}\label{eq:SMScatterOff}
    \vect{S}_j = \left(\begin{array}{cc} t&c_j\\ c_j&t \end{array}\right);
\end{equation}
the transmission and conversion coefficients can be written as
\begin{equation}\label{eq:SMCoeff}
    t = \frac{\kappa+2i\omega}{\kappa-2i\omega},\qquad 
    c_j = -\frac{8 g_{1j}g_{2j}\kappa}{(\kappa-2i\omega)^2(\gamma-2i\omega)}.
\end{equation}
In the transmission coefficients, we dropped the effect of the opto- and electromechanical interaction, which is, for off-resonant signals ($g_i<\kappa$, $\omega\sim\kappa$), small compared to the direct transmission.
Eqs. \eqref{eq:SMCoeff} do not hold on resonance and thus cannot give us the proper spectrum;
when estimating the bandwidth of large arrays, we are, however, interested only in frequencies far off resonance where the approximation given by Eq. \eqref{eq:SMCoeff} is valid.

Although the transfer matrices differ from site to site, they can all be diagonalized simultaneously.
Using the transformation $\vect{U}^{-1}\vect{S}_j\vect{U}$ with
\[
    \vect{U} = \frac{1}{\sqrt{2}}\left(\begin{array}{cc} 1&1 \\ 1&-1 \end{array}\right),
\]
we obtain the diagonal form
\begin{equation}
    \vect{S}_j^\diag = \diag(t+c_j,t-c_j).
\end{equation}
The transfer matrix of the array is, in the diagonal form, given by the product of transfer matrices of the individual transducers,
\begin{equation}
    \vect{T}_\diag = \prod_{j=1}^N \vect{S}_j^\diag = \diag\left(\prod_{j=1}^N(t+c_j), \prod_{j=1}^N(t-c_j)\right).
\end{equation}
For weak coupling, $c_j\ll 1$ far off resonance, we can keep only terms linear in $c_j$,
\begin{equation}
    \prod_{j=1}^N(t\pm c_j) \approx t^N \pm t^{N-1}\sum_{j=1}^N c_j.
\end{equation}
The conversion coefficient $T_{21}$ of the array can be found by transforming $\vect{T}_\diag$ back to the lab frame, $\vect{T} = \vect{UT}_\diag\vect{U}^{-1}$, which yields
\begin{equation}\label{eq:SMconversion}
    T_{21} = t^{N-1}\sum_{j=1}^N c_j.
\end{equation}

Next, we assume that the coupling rates are varied linearly across the array, $g_{1j} = j g/N$, $g_{2j} = g(1-j/N)$.
[In the numerical simulations, we use tanh variation of the coupling; see Eq.~\eqref{eq:GOpt}.
Nevertheless, the linear variation enables us to find an analytical expression for the conversion bandwidth.]
We can now perform the sum in Eq.~\eqref{eq:SMconversion} and obtain the conversion coefficient
\begin{equation}
    T_{21}(\omega) = \left(\frac{\kappa+2i\omega}{\kappa-2i\omega}\right)^2\frac{8g^2\kappa}{(\kappa-2i\omega)^2(\gamma-2i\omega)}\frac{1-N^2}{6N}.
\end{equation}
To find the bandwidth, we put $|T_{21}(\omega)|^2 = \frac{1}{2}$ and solve for frequency.
We obtain a cubic equation in $\omega^2$ with two complex roots; from the third, real root, we get
\begin{equation}\label{eq:SMfreq}
    \omega_\pm = \pm\frac{-3^{2/3}\kappa^2 + 3^{1/3}\left(6\sqrt{2}g^2\kappa\displaystyle\frac{N^2-1}{N} + \kappa\sqrt{72g^4\frac{(N^2-1)^2}{N^2} + 3\kappa^4}\right)^{2/3}}{6\left(6\sqrt{2}g^2\kappa\displaystyle\frac{N^2-1}{N} + \kappa\sqrt{72g^4\frac{(N^2-1)^2}{N^2} + 3\kappa^4}\right)^{1/3}}.
\end{equation}
The frequency $\omega_+$ is always positive (and $\omega_-$ is always negative); the negative term in the numerator is compensated by the last term under the square root in the numerator.
The conversion bandwidth is $\Delta\omega = \omega_+ - \omega_- = 2\omega_+$.
In the large-array limit, $g\sqrt{N} > \kappa$, the bandwidth can be further simplified to
\begin{equation}
    \Delta\omega = \left(\frac{4\sqrt{2}}{3} g^2\kappa N\right)^{1/3}.
\end{equation}
For symmetric transducer arrays, the bandwidth depends only on the maximum coupling rate, the cavity linewidth, and the array size.

\subsection{Unequal cavity decay rates}

When the decay rates of the microwave and optical fields differ, $\kappa_1\neq\kappa_2$, the two propagating fields have different dispersion;
the mechanically dark mode does not exist and the conversion bandwidth saturates.
To see this, we consider a single mode waveguide of length $L$ unidirectionally coupled to $N$ identical cavities;
with linear dispersion of the waveguides and frequency rescaled to the resonant frequency of the cavities, we get the Hamiltonian
\begin{equation}
    H = -iv\int_0^L\dd{}z\,a^\dagger (z)\frac{\partial}{\partial z} a(z) + g \sum_{j=1}^N a^\dagger (z_j) c_j + c_j^\dagger a(z_j).
\end{equation}
We now represent the $N$ cavities as a continuum along $z$ by introducing the continuous operator
\begin{equation}
    \tilde{c}(z,t) = \frac{1}{\sqrt{n}} \sum_j \delta(z-z_j)c_j(t),
\end{equation}
where $n=N/L$ is the density of cavities along the waveguide, and $\tilde{c}(z)$ satisfies the usual commutation relations.
The dynamics can be diagonalized in frequency space, i.e., by writing $a(z,t) = \sum_ka_k e^{i (kz - \omega t)}/\sqrt{L}$, and $\tilde{c}(z,t) = \sum_k \tilde{c}_k e^{i (kz - \omega t)}/\sqrt{L}$, to obtain the dispersion relation
\begin{equation}
\label{eq:dispersion}
    k = \frac{\omega}{v} - \frac{\kappa^2}{v \, \omega},
\end{equation}
where $\kappa = \sqrt{n} g$. The eigenoperators are $d_k = (\omega a_k + {\kappa}\tilde{c}_k)/\sqrt{\omega^2 + \kappa^2}$ and the Hamiltonian can be written as $H = \sum_k \omega_k d_k^\dagger d_k$,
where $\omega_k$ is obtained from the dispersion relation \eqref{eq:dispersion}.

The transducer array contains two such waveguides, each coupled to a continuum of mechanical oscillators with rates $g_{1,k}$ and $g_{2,k}$ through a beam splitter interaction.
Its Hamiltonian is
\begin{equation}
    H = \sum_k \omega_{1,k} d_{1,k}^\dagger d_{1,k} + \sum_k \omega_{2,k} d_{2,k}^\dagger d_{2,k} + \sum_k {\omega}_{m,k} b_k^\dagger b_{k} + \sum_k(g_{1,k}d_{1,k} + g_{2,k}d_{2,k}) b_k^\dagger + \Hc, 
\end{equation}
where $\omega_{1,k}$ and $\omega_{2,k}$ are the dispersion curves of the two cavity-waveguide continua with cavity-waveguide coupling $\kappa_1$ and $\kappa_2$ respectively. The continuum of mechanical oscillators has the annihilation operators  $\hat{b}_k$ and dispersion ${\omega}_{m,k}$. 
As before the equations of motion for this Hamiltonian are linear and can be diagonalized  in Fourier space.
In the special case of $\kappa_1=\kappa_2$, which implies $\omega_{1,k} = \omega_{2,k}$, this system of equations has the mechanically dark state $d_\mathrm{dark} \propto g_{2,k}d_{1,k} - g_{1,k}d_{2,k}$ as its eigenvector.
If $\omega_{1,k} \neq \omega_{2,k}$ the dark state no longer exists; the conversion bandwidth, therefore, saturates.

Interestingly, the cubic-root scaling can be recovered if we vary the cavity linewidths across the array; cf. Fig.~\ref{fig:Imbalance}(a).
We vary the decay rates in the same direction as the coupling rates---the microwave cavity linewidth $\kappa_1$ increases while the linewidth of the optical cavities $\kappa_2$ decreases---and
observe that, as long as the linewidths are equal at one site in the array, the bandwidth monotonically increases with array size; see Fig.~\ref{fig:Imbalance}(b).
This variation can be moderate and does not require the decay rate to approach zero at the ends of the array.

\begin{figure}[t!]
    \centering
    \includegraphics[width=0.55\linewidth]{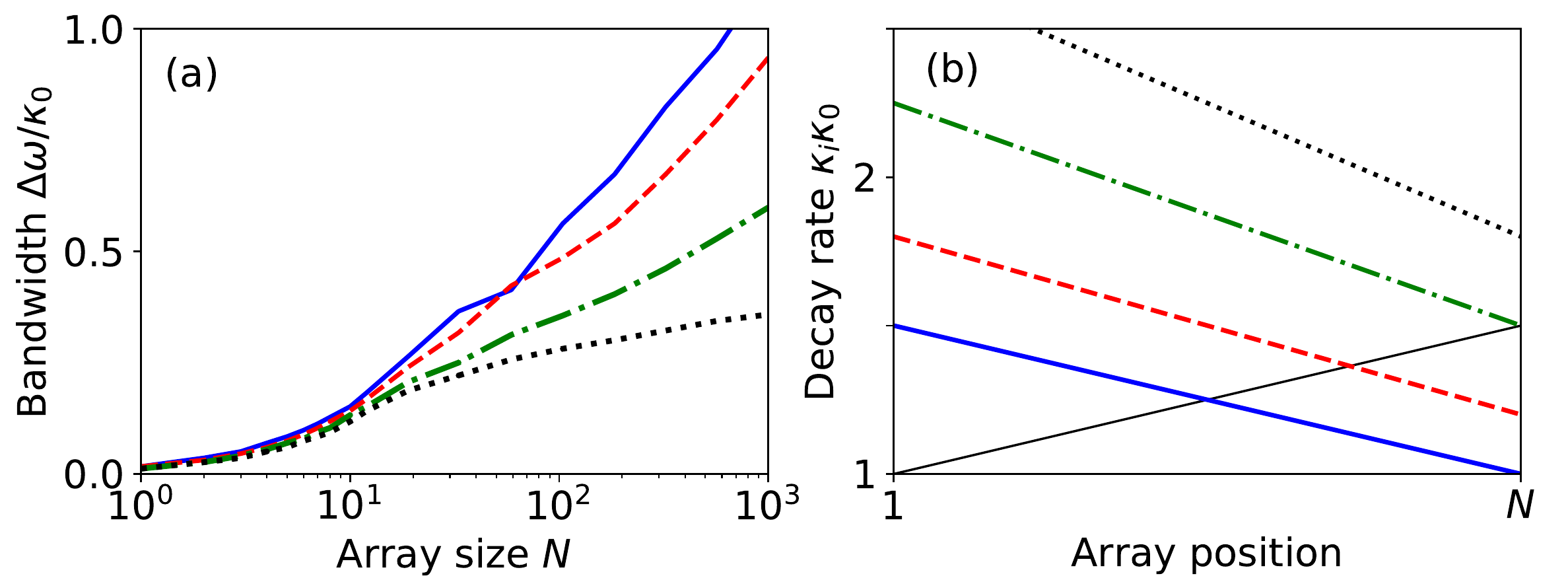}
    \caption{\label{fig:Imbalance}
        (a) Bandwidth with variable cavity decay rates.
        Throughout the array, the microwave decay rate changes from $\kappa_1 (0) = \kappa_0$ to $\kappa_1(L) = 1.5\kappa_0$; the optical linewidth varies in the opposite direction from $\kappa_2(0) = 1.5\kappa_f$ to $\kappa_2(L) = \kappa_f$.
        The curves show bandwidth for $\kappa_f = \kappa_0$ (solid blue line), $\kappa_f = 1.2\kappa_0$ (dashed red line), $\kappa_f = 1.5\kappa_0$ (dot-dashed green line), and $\kappa_f = 1.8\kappa_0$ (dotted black line).
        For the coupling, we have $\bar{g}_1 = 0.08/\kappa_0$.
        (b) Variation of the decay rates used to obtain the results in panel (a).
        For all four curves, the microwave decay rate varies from $\kappa_1(0) = \kappa_0$ to $\kappa_1(L) = 1.5\kappa_0$ (thin black line).
        The optical decay rates vary in the opposite direction; the color coding corresponds to the data shown in (a).
    }
\end{figure}

\section{Frequency conversion in small arrays}

For small transducer arrays, the conversion bandwidth is smaller than the cavity linewidth and the cavity fields can thus be adiabatically eliminated from the dynamics.
This enables us to obtain a simplified expression for the scattering matrix where only the effective coupling rates $\Gamma_i = g_i^2/\kappa_i$ of the propagating fields to the mechanical oscillators are relevant.
Furthermore, owing to the small parameter space, we can fully optimize the array for maximum bandwidth.

Starting from the Langevin equations \eqref{eq:SLangevin} and adiabatically eliminating the cavity fields from the dynamics, we obtain the scattering matrix
\begin{equation}\label{eq:scattering}
    \vect{S}(\omega) = \frac{1}{2(\Gamma_1+\Gamma_2)-i\omega}\left(\begin{array}{cc}
        -2(\Gamma_1-\Gamma_2) -i\omega & -4\sqrt{\Gamma_1\Gamma_2} \\
        -4\sqrt{\Gamma_1\Gamma_2} & 2(\Gamma_1-\Gamma_2) -i\omega
    \end{array}\right);
\end{equation}
here, we neglect mechanical dissipation for simplicity.
The scattering properties of a single transducer are fully determined by the effective coupling rates $\Gamma_i = g_i^2/\kappa_i$ and by the Fourier frequency $\omega$.
In the following, we will assume that the sum of the two effective coupling rates is constant across the array, $\Gamma_1^j + \Gamma_2^j = \Gamma = \const$
The transducer array is thus characterized by the total coupling $\Gamma$ and each transducer by one of the coupling rates, say, $\Gamma_1$. 
Our goal is now to maximize the conversion bandwidth of an array of $N$ transducers.

We will, furthermore, focus only on symmetric arrays; that is, arrays that are invariant under mirroring (i.e., the exchange $1\to N$, $2\to N-1$, etc.) and permutation of modes.
For such an array, we can collect the coupling rates $\Gamma_1^j$ in a vector $\gvec{\Gamma}_1 = (\Gamma_1^1,\Gamma_1^2,\ldots,\Gamma-\Gamma_1^2,\Gamma-\Gamma_1^1)^T$. 
Numerical simulations (for $N\leq 3$) indicate that such arrays are optimal for maximizing the conversion bandwidth and this assumption allows us to halve the number of parameters we have to optimize over, significantly simplifying the numerical optimization.

A simulation for $N = 2$ reveals a problem (see Fig.~\ref{fig:N2}): 
When the overall conversion bandwidth is maximized, a dip appears in the middle of the conversion spectrum. 
Quantifying the conversion solely on the basis of the overall bandwidth would therefore be misguided. 
There is an apparent tradeoff to be made: 
We cannot improve the bandwidth without sacrificing conversion efficiency on resonance.
Similar behaviour can be observed also for larger arrays. 
In general, the conversion spectrum is formed by $N$ peaks; only when these peaks are closely spaced can we obtain a conversion spectrum with a flat (or an almost flat) top.

\begin{figure}
    \centering
    \includegraphics[width=0.67\linewidth]{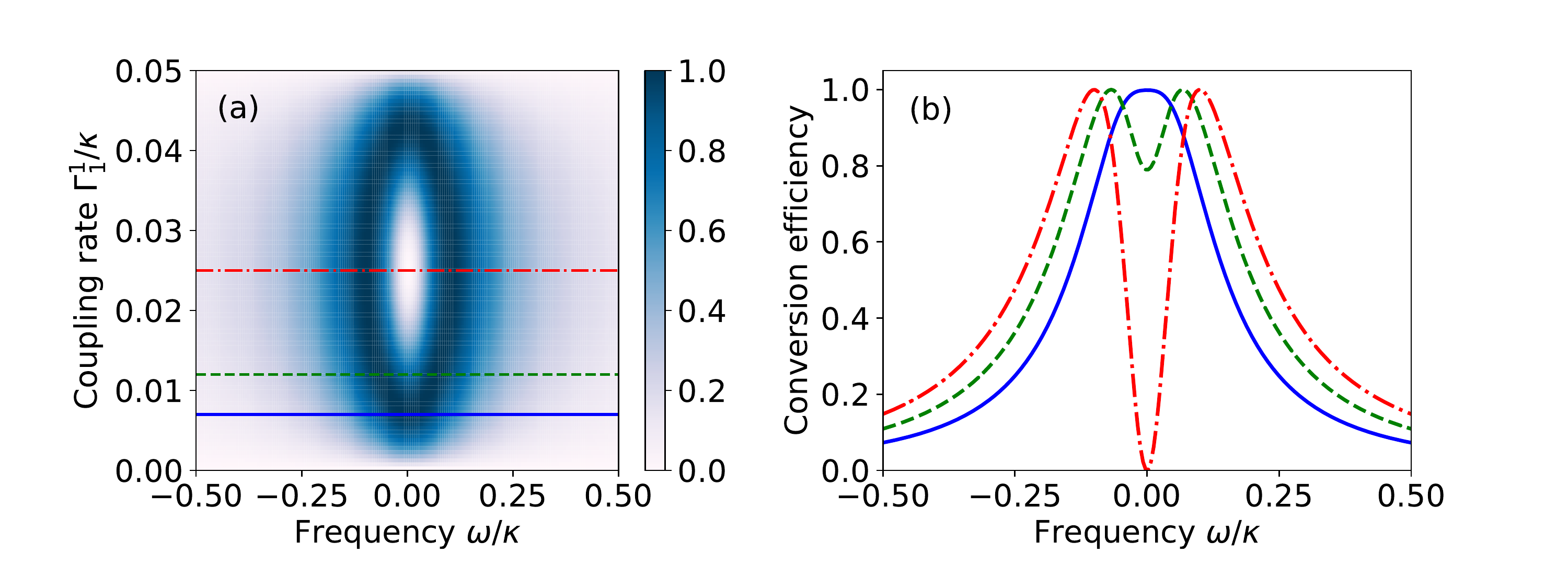}
    \caption{\label{fig:N2} (a) Conversion efficiency as a function of frequency and coupling rate $\Gamma_1^1$ for a two-transducer array.
    (b) Conversion spectra for arrays with $\Gamma_1^1/\kappa = 0.007, 0.012, 0.025$ (solid blue, dashed green, and dot-dashed red line, respectively).
    The three curves correspond to the horizontal lines in panel (a).
    For both panels, we have $\Gamma/\kappa = 0.05$.}
\end{figure}

This problem can be resolved by optimizing the bandwidth with a constraint on the depth of the local minimum of conversion efficiency. 
We say that we require the conversion to work with minimum efficiency of, say, 99 \% and ask what bandwidth we can reach with this constraint. 
In the case of $N = 2$ transducers, the bandwidth is about $0.33\kappa$ with $\Gamma_1^1/\kappa = 0.008$ and $\Gamma/\kappa = 0.05$ (for a single transducer, the bandwidth would be $0.2\kappa$).
Full results of the constrained optimization are shown in Fig.~\ref{fig:Optimization}(a,b). 
In panel (a), scaling of the bandwidth with the array size is plotted and two values of allowed minima are considered, namely 0.9 (blue squares) and 0.99 (green circles). 
In both cases, the scaling of bandwdith is close to linear in the array size (shown as the red line).
In panel (b), the bandwidth is plotted as a function of the minimum allowed efficiency for an array of $N=6$ transducers.

\begin{figure}[b!]
    \centering
    \includegraphics[width=0.78\textwidth]{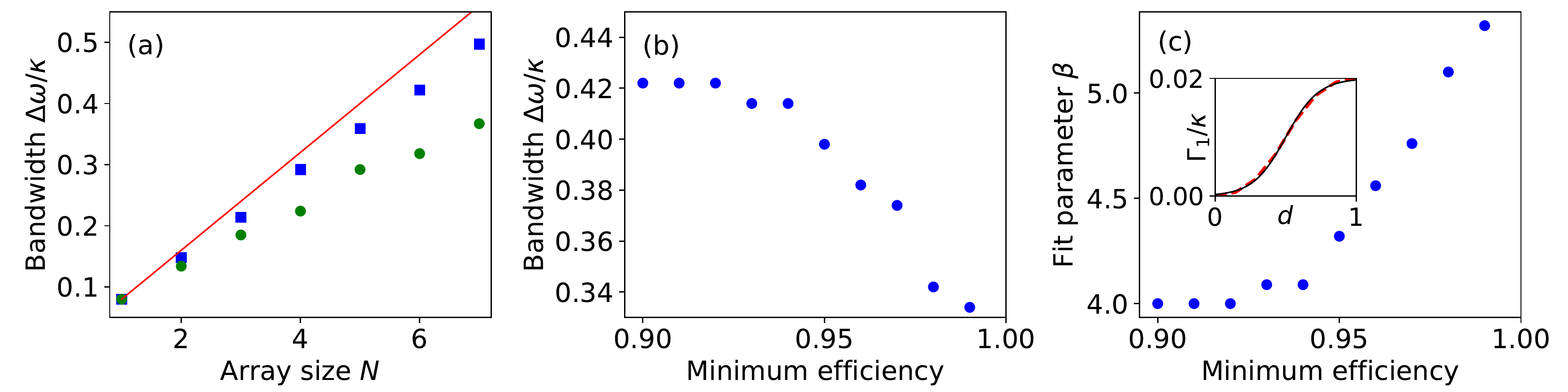}
    \caption{\label{fig:Optimization}
    (a) Conversion bandwidth versus array size $N$ for minimum efficiency 0.9 (blue squares) and 0.99 (green circles).
    The thin red line is the linear fit $4\Gamma N$ and the overall coupling strength is $\Gamma/\kappa = 0.02$.
    (b) Bandwidth for an array with $N=6$ transducers as a function of the minimum allowed efficiency.
    (c) The fitting parameter $\beta$ for the data shown in (b).
    The inset shows the numerically optimized coupling rates (dashed red line) and the fit (thin black line) for minimum efficiency of 0.95.
    }
\end{figure}

The optimum coupling $\Gamma_1$ can be approximated with the fit
\begin{equation}\label{eq:GOpt}
    \Gamma_{1,\mathrm{opt}}(d) = \frac{\Gamma}{2}\left\{\tanh\left[\beta\left(d-\frac{1}{2}\right)\right]+1\right\}.
\end{equation}
The fitting parameter $\beta$ is, to a good approximation, independent of the array size if we include two more transducers: one, with $\Gamma_1 = 0$, for position $j = 0$ and another, with $\Gamma_1 = \Gamma$, for $j = N+1$.
With this extension (which does not change the conversion efficiency or bandwidth), the normalized array position $d = j/(N+1)\in[0,1]$.
The fitting parameter, for various values of the minimum conversion efficiency, is shown in Fig~\ref{fig:Optimization}(c);
the inset shows the coupling rates [obtained by numerical optimization and from the fit~\eqref{eq:GOpt}] for minimum efficiency of 0.95.
In the main text, we use $\beta = 4.5$, corresponding to minimum efficiency of about 95 \%.

\section{Mechanical noise}

The scattering matrix \eqref{eq:SMScattering} [or \eqref{eq:SMScatterRes}] takes into account the loss of signal through mechanical decay but not the associated noise.
To analyze this noise, we can introduce a source term into the scattering process;
we start from the Heisenberg-Langevin equations
\begin{subequations}\label{eq:SHLNoise}
\begin{align}
    \dot{c}_i &= -\frac{\kappa_i}{2}c_1 - ig_ib + \sqrt{\kappa_i}a_i(z_j^-),\\
    \dot{b} &= -\frac{\gamma}{2}b-ig_1c_1-ig_2c_2 + \sqrt{\gamma}f_j.
\end{align}
\end{subequations}
Eqs.~\eqref{eq:SHLNoise} are, up to the mechanical noise (the last term in the last equation, $\sqrt{\gamma}f_j$, where $f_j$ is the noise operator describing the reservoir) identical to Eqs.~\eqref{eq:SLangevinHEq}.

In matrix form, we can write
\begin{subequations}
\begin{align}
    \dot{\vect{a}} &= \vect{Aa} + \vect{Ba}_\mathrm{in} +\vect{E}f_j \\
    \vect{a}_\mathrm{out} &= \vect{Ca} + \vect{Da}_\mathrm{in},
\end{align}
\end{subequations}
where the matrices $\vect{A}$, $\vect{B}$, $\vect{C}$, $\vect{D}$ are the same as previously and $\vect{E} = (0,0,\sqrt{\gamma})^T$.
The solution in terms of the state-space model is formally analogous to the previous case;
we can write
\begin{align}
    \vect{a}(z_j^+,\omega) &= [\vect{D}+\vect{C}(\vect{A}+i\omega\vect{I}_3)^{-1}\vect{B}]\vect{a}(z_j^-,\omega) + \vect{C}(\vect{A}+i\omega\vect{I}_3)^{-1}\vect{E}f_j(\omega) \nonumber\\
    &= \vect{S}_j(\omega)\vect{a}(z_j^-,\omega) + \vect{V}_j(\omega)f_j.
\end{align}
Here, $\vect{V}_j(\omega)$ describes the coupling of the bath to the propagating fields.
For a symmetric transducer ($\kappa_1 = \kappa_2 = \kappa$, $\bar{g}_1 = \bar{g}_2 = g$) with linear variation of coupling rates, we have
\begin{equation}\label{eq:SMNoise}
    \vect{V}_j(\omega) = -\frac{4ig\sqrt{\kappa\gamma}}{4(g/N)^2(N^2-2jN+2j^2)+(\kappa-2i\omega)(\gamma-2i\omega)}\left(\frac{j}{N},1-\frac{j}{N}\right)^T.
\end{equation}

We can obtain the total added noise by incoherently summing the noise contributions from each transducer in the array,
\begin{equation}\label{eq:SMNoiseAdd}
    S_\mathrm{add}^2(\omega) = \sum_{j=1}^N|\gvec{\chi}_j(\omega)|^2S_f^2(\omega);
\end{equation}
here, $S_f^2(\omega) = 2\nbar +1$ is the noise spectral density of the thermal force $f_j$ (we assume that all mechanical reservoirs have the same temperature).
Moreover,
\begin{equation}
    \gvec{\chi}_j(\omega) = \prod_{k=j+1}^N\vect{S}_k(\omega)\vect{V}_j(\omega)
\end{equation}
is the noise susceptibility of the $j$th transducer.
In Eq.~\eqref{eq:SMNoiseAdd}, the absolute value of the noise susceptibility is taken elementwise,
$|\gvec{\chi}_j(\omega)|^2 = [|\chi_{1j}(\omega)|^2, |\chi_{2j}(\omega)|^2]^T$;
we can thus decribe noise added to both output modes.

In the following, we will consider two different regimes for evaluating the added noise for large arrays:
In the first one, we operate close to resonance and use the scattering matrix given by Eq. \eqref{eq:SMScatterRes};
alternatively, we consider off-resonant signals and use approach similar to the one employed to find the conversion bandwidth.
Finally, we evaluate the added noise for small arrays numerically and show that it is suppressed compared to conversion using a single transducer.

\subsection{Added noise on resonance}
On cavity resonance, the scattering matrix of a single transducer is real and given by Eq. \eqref{eq:SMScatterRes}.
It can be diagonalized by the orthogonal transformation $\vect{O}_j\vect{S}_j\vect{O}_j^T$ with
\begin{equation}
    \vect{O}_j = \frac{1}{\sqrt{\tilde C_1+\tilde C_2}}\left(\begin{array}{cc}
        \sqrt{\tilde C_1} & \sqrt{\tilde C_2} \\ -\sqrt{\tilde C_2} & \sqrt{\tilde C_1}
    \end{array}\right);
\end{equation}
with strong total cooperativity, $\tilde C_1 + \tilde C_2 > 1$, the diagonal form of the scattering matrix is $\vect{S}_j^\diag = \diag(-1,1)$.
The matrix $\vect{O}_j$ describes a rotation with rotation angle $\tan\theta_j = \sqrt{\tilde C_2/\tilde C_1}$.
In a large array, the rotation angles for two neighboring transducers are almost identical;
we therefore have $\vect{O}_{j+1}^T\vect{O}_j = \vect{I}_2 + \gvec{\epsilon}$ with
\begin{equation}
    \gvec{\epsilon} = \frac{1}{N}\left(\begin{array}{cc} 0&1 \\ -1&0 \end{array}\right).
\end{equation}

When evaluating the noise susceptibility, we can diagonalize all scattering matrices simultaneously;
to first order in the small correction $\gvec{\epsilon}$, we obtain the expression
\begin{align}
    \gvec{\chi}_j(0) &= \prod_{k=j+1}^N\vect{S}_k(0)\vect{V}_j(0) 
        = \vect{O}_N^T\left(\prod_{k=j+1}^N\vect{S}_j^\diag + \sum_{k=j+1}^{N-1}\prod_{m=k+1}^N\vect{S}_m^\diag\gvec{\epsilon}\prod_{n=j+1}^k\vect{S}_n^\diag\right)\vect{O}_{j+1}\vect{V}_j(0) \nonumber\\
        &= -2i\frac{\sqrt{\tilde C}}{\tilde C+1}\left(
        (-1)^{N+j}, 
        \frac{1}{2N}[-1+(-1)^{N+j}]
        \right)^T
\end{align}
with $\tilde C = 4g^2/\kappa\gamma$.
For the total added noise, we now have
\begin{equation}
    S_\mathrm{add}^2(0) = \frac{4\tilde C(2\nbar+1)}{(\tilde C+1)^2}\left(N,\frac{1}{2N}\right)^T.
\end{equation}
This expression clearly reveals the advantage of using the dark mode for frequency conversion:
With the dark mode, the noise is suppressed for $\tilde CN\gg\nbar$ (second component of the added noise) and larger array thus helps to reduce the noise.
The noise in the bright mode, on the other hand, grows with array size; we need $\tilde C\gg\nbar N$.

\subsection{Off-resonant noise}
In estimating the added noise off resonance, we proceed similar to evaluating the conversion bandwidth.
We assume that the scattering matrix of a single transducer is given by Eq. \eqref{eq:SMScatterOff} which enables us to simultaneously diagonalize the scattering matrices of all transducers.
We then transform the added noise $\vect{V}_j(\omega)f_j$ by the same transformation and thus obtain the total added noise.

We do not reproduce the whole derivation here since the resulting expression is too cumbersome.
The result is the same as for noise on resonance, namely, that
thermal noise is suppressed for $\tilde CN\gg\nbar$.
This result can be easily understood:
The conversion proceeds via the mechanically dark mode of the two propagating fields and thus is not directly affected by mechanical losses and noise;
thermal noise comes only from the crosstalk between the two normal modes.
The added noise can thus be limited by reducing either the cross talk (by improving the adiabaticity of the process, which we can achieve by increasing the array size $N$)
or the coupling of the bright mode to the mechanical bath (by increasing the cooperativity).

\subsection{Mechanical noise in small arrays}

\begin{figure}
    \centering
    \includegraphics[width=0.85\textwidth]{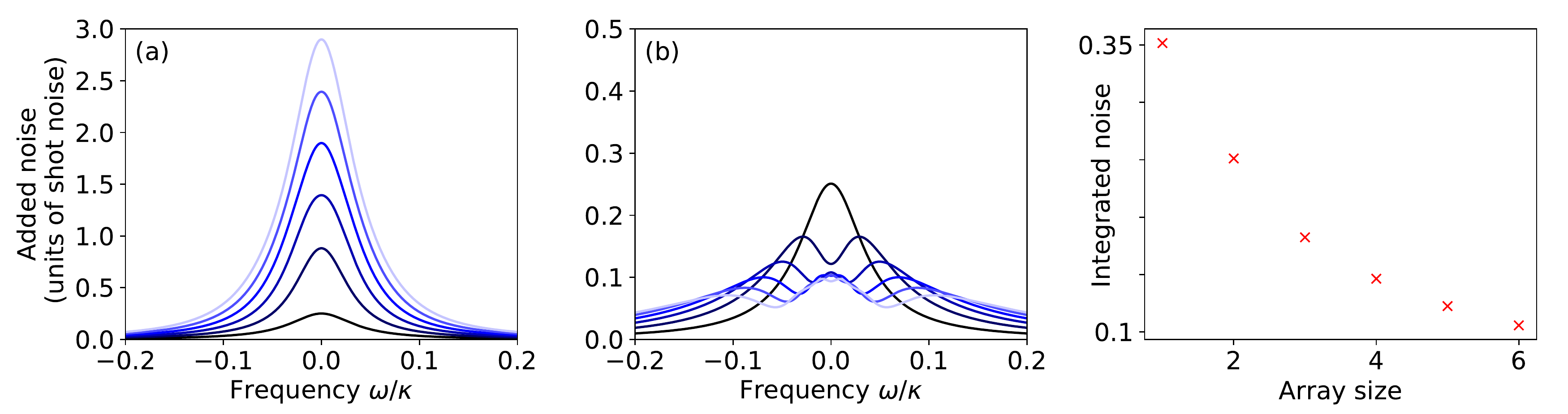}
    \caption{\label{fig:Noise}
    Spectral density of the added noise with increasing array size for frequency conversion via the bright (a) and dark mode (b).
    The black lines correspond to a single transducer; towards light colors, the array size increases up to $N = 6$.
    Panel (c) shows the total added noise integrated over the conversion spectrum as a function of the array size for the spectral density shown in (b).
    The parameters used are $\Gamma = 0.02\kappa$ (corresponding to coupling $g = 0.1\kappa$, mechanical oscillator of frequency $\omega_m = 10\kappa$ and quality factor $Q_m =\omega_m/\gamma = 2\times 10^5$ and average thermal occupation $\nbar = 100$; the optomechanical cooperativity $C = 2\Gamma/\gamma\nbar = 8$.}
\end{figure}

The scaling of added thermal noise with collective optomechanical cooperativity is valid only in the adiabatic limit $g\sqrt{N}>\kappa$.
To analyze how conversion in small transducer arrays is affected by added noise, we evaluate the spectral density \eqref{eq:SMNoiseAdd} numerically and plot the results in Fig.~\ref{fig:Noise}.
Panel (a) shows that frequency conversion via the mechanically bright mode [first component of the noise spectral density \eqref{eq:SMNoiseAdd}] leads to increased noise for larger arrays.
On the other hand, conversion via the mechanically dark mode [panel (b)] leads to suppression of thermal noise close to resonance.
The effect is not as pronounced as for large arrays where the adiabatic condition is fulfilled;
in fact, the total added noise (obtained by integrating the noise spectral density over the conversion spectrum) slightly increases with array size [panel (c)].
This increase is, however, sublinear in the array size, indicating that the use of the dark mode helps to suppress the thermal mechanical noise.

\section{Stokes scattering noise}

The effect of counterrotating terms can be obtained by a modification of the state space model that includes the creation operators.
We start from the full linearized Hamiltonian for a single transducer in the lab frame
\begin{equation}
    H = \omega_m(c_1^\dagger c_1 + c_2^\dagger c_2 + b^\dagger b) + g_1(c_1+c_1^\dagger)(b+b^\dagger) + g_2(c_2 + c_2^\dagger)(b+b^\dagger).
\end{equation}
Previously, we applied the rotating wave approximation to get rid of the terms with $c_i b$ and $c_i^\dagger b^\dagger$
which is justified for $\kappa_i\ll\omega_m$.
In the following, we will keep these terms in the interaction and study the resulting noise added to the converted signal.

We write the Langevin equations in the matrix form
\begin{subequations}
\begin{align}
    \dot{\vect{a}} &= \vect{Aa} + \vect{Ba}_\mathrm{in},\\
    \vect{a}_\mathrm{out} &= \vect{Ca} + \vect{Da}_\mathrm{in}.
\end{align}
\end{subequations}
Here, we use the definitions $\vect{a} = (c_1,c_2,b,c_1^\dagger,c_2^\dagger,b^\dagger)$, $\vect{a}_\mathrm{in} = [a_1(z_j^-),a_2(z_j^-),a_1^\dagger(z_j^-),a_2^\dagger(z_j^-)]^T$, and similar for $\vect{a}_\mathrm{out}$;
additionally, we have the matrices
\begin{subequations}
\begin{align}
    \vect{A} &= \left(\begin{array}{cccccc}
        -i\omega_m-\frac{\kappa_1}{2} & 0 & -ig_1 & 0 & 0 & -ig_1 \\
        0 & -i\omega_m -\frac{\kappa_2}{2} & -ig_2 & 0 & 0 & -ig_2 \\
        -ig_1 & -ig_2 & -i\omega_m-\frac{\gamma}{2} & -ig_1 & -ig_2 & 0 \\
        0 & 0 & ig_1 & i\omega_m - \frac{\kappa_1}{2} & 0 & ig_1 \\
        0 & 0 & ig_2 & 0 & i\omega_m - \frac{\kappa_2}{2} & ig_2 \\
        ig_1 & ig_2 & 0 & ig_1 & ig_2 & i\omega_m-\frac{\gamma}{2}
    \end{array}\right),\\
    \vect{B} &= \left(\begin{array}{cccc}
        \sqrt{\kappa_1} &0&0&0 \\ 0&\sqrt{\kappa_2}&0&0 \\ 0&0&0&0 \\ 0&0&\sqrt{\kappa_1}&0 \\ 0&0&0&\sqrt{\kappa_2} \\ 0&0&0&0
    \end{array}\right),\\
    \vect{C} &= \vect{B}^T,\\
    \vect{D} &= -\vect{1}_4.
\end{align}
\end{subequations}
The action of a transducer can thus be described by the scattering matrix
\begin{equation}
    \vect{a}_\mathrm{out}(\omega) = \vect{S}(\omega)\vect{a}_\mathrm{in}(\omega) = [\vect{D}-\vect{C}(\vect{A}+i\omega\vect{1}_6)^{-1}\vect{B}]\vect{a}_\mathrm{in}(\omega)
\end{equation}
and the effect of the whole array is obtained by multiplying the scattering matrices of individual transducers.
Note that this scattering matrix describes a general Bogoliubov transformation mixing creation and annihilation operators.
The scattering process therefore does not conserve the number of incoming and outgoing photons.
In the following we will consider the outgoing excess  photons as added noise.  

When we are using the scattering matrix formalism to describe conversion of a signal initially in the mode $a_1(0)$ to the mode $a_2(L)$, the corresponding output is characterized by the following relation:
\begin{equation}
    a_2(L,\omega) = T_{21}(\omega)a_1(0,\omega) + T_{22}(\omega)a_2(0,\omega) + T_{23}(\omega)a_1^\dagger(0,\omega) + T_{24}(\omega)a_2^\dagger(0,\omega).
\end{equation}
Here, the first term on the RHS describes the conversion and the second term direct transmission; we already understand their behavior.
(Note, however, that since we now operate in the lab frame, the conversion is peaked around $\omega = \omega_m$.)
The remaining two terms---characterized by the elements $T_{23}$ and $T_{24}$---describe effective two-mode and one-mode squeezing processes, respectively.
The average occupation of the mode $a_2(L)$ for each of these processes is given by the modulus squared of the corresponding element of the scattering matrix, $|T_{23}(\omega)|^2$, $|T_{24}(\omega)|^2$, which thus quantify the noise added at a given frequency $\omega$.

Since the resulting expressions are too cumbersome to allow direct physical insight, we study the added noise numerically in Fig.~\ref{fig:Counterrotating},
where we study the number of added photons integrated over the conversion bandwidth,
\begin{equation}
    n_\mathrm{Stokes} = \int_{\omega_m - \Delta\omega/2}^{\omega_m + \Delta\omega/2}d\omega [|T_{23}(\omega)|^2 + |T_{24}(\omega)|^2],
\end{equation}
as a function of the sideband ratio $\kappa/\omega_m$ (a) and array size $N$ (b).
The variation of the total added noise with the sideband ratio is faster than quadratic (which is the scaling one might expect since the problem is similar to the backaction limit of sideband cooling) since variation of the cavity linewidth (used in the simulations to change the sideband ratio) modifies also the conversion bandwidth and thus the integration region.
Secondly, the added noise scales exponentially with array size for large arrays, which prohibits their use for frequency conversion.
With moderate-sized arrays (with up to $\sim 80$ transducers with the parameters used here), however, the total noise is comparable to, or even smaller than with, a single transducer.
Additionally, the added noise is concentrated into a narrow spectral region around the sideband, as illustrated in Fig.~\ref{fig:Counterrotating}(c).

\begin{figure}
    \centering
    \includegraphics[width=0.85\linewidth]{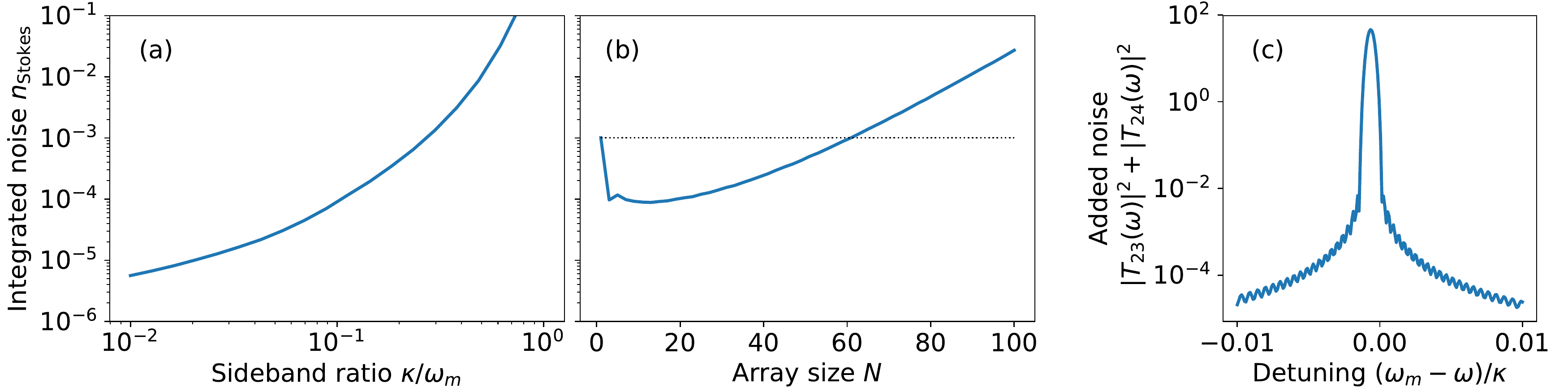}
    \caption{\label{fig:Counterrotating}
    Total integrated noise $n_\mathrm{Stokes}$ as a function of (a) the sideband ratio $\kappa/\omega_m$ for $N = 10$ transducers and (b) the array size $N$ for the sideband ratio $\kappa/\omega_m = 0.1$.
    (c) Spectrum of the added noise $|T_{23}(\omega)|^2 + |T_{24}(\omega)|^2$ for $N = 100$ and $\kappa/\omega_m = 0.1$.
    The horizontal line in panel (b) shows the photons added for a single transducer.}
\end{figure}

\section{Optical losses}

To model optical and microwave losses, we modify the state-space model describing our transducers.
In the Heisenberg-Langevin equations for a single transducer, the cavity modes decay through two channels, giving rise to left- and right-propagating fields,
\begin{subequations}
\begin{align}
    \dot{c}_i &= - \frac{\kappa_{i,R}+\kappa_{i,L}+\kappa_{i,\mathrm{int}}}{2}a_i - ig_ib + \sqrt{\kappa_{i,R}}a_{i,R}^\mathrm{in}+\sqrt{\kappa_{i,L}}a_{i,L}^\mathrm{in} + \sqrt{\kappa_{i,\mathrm{int}}}a_{i,\mathrm{int}}^\mathrm{in}, \\
    \dot{b} &= - \frac{\gamma}{2}b - ig_1a_1 - ig_2a_2 + \sqrt{\gamma}b_\mathrm{in}.
\end{align}
\end{subequations}
The right-propagating fields describe the signal; the left-propagating fields represent unwanted backscattering.
We also included intrinsic cavity loss at rate $\kappa_{i,\mathrm{int}}$ and the associated noise operator $a_{i,\mathrm{int}}^\mathrm{in}$;
in the previous analysis, we had $\kappa_{i,L} = \kappa_{i,\mathrm{int}} = 0$.

To solve the state-space model, we use the input-output relations
\begin{equation}
    a_{i,\alpha}^\mathrm{out} = \sqrt{\kappa_{i,\alpha}}c_i - a_{i,\alpha}^\mathrm{in}
\end{equation}
with $\alpha\in\{R,L\}$.
Next, we collect the localized modes in the vector $\vect{a} = (c_1, c_2, b)^T$ and the propagating fields in the vectors
$\vect{a}_\mathrm{in,out} = (a_{1,R}^\mathrm{in,out}, a_{2,R}^\mathrm{in,out}, a_{1,L}^\mathrm{in,out}, a_{2,L}^\mathrm{in,out})^T$;
we group the propagating fields by their direction of propagation.
We can now write the scattering matrix in the block form
\begin{equation}
    \vect{S} = \left(\begin{array}{cc} \vect{S}_R & \vect{S}_{RL} \\ \vect{S}_{LR} & \vect{S}_L \end{array} \right).
\end{equation}
Here, the diagonal elements describe scattering processes, in which the field does not change direction of propagation;
the off-diagonal terms describe scattering processes, in which the propagation direction is changed.

The scattering matrix describes the relation between input and output fields.
To convert it into a transfer matrix (which describes the relation between fields at positions $z_1$ and $z_2$),
we rewrite the input and output vectors in position coordinates,
\begin{subequations}
\begin{align}
    \vect{a}_\mathrm{in} &= [a_{1,R}(z^-), a_{2,R}(z^-), a_{1,L}(z^+), a_{2,L}(z^+)]^T,\\
    \vect{a}_\mathrm{out} &= [a_{1,R}(z^+), a_{2,R}(z^+), a_{1,L}(z^-), a_{2,L}(z^-)]^T.
\end{align}
\end{subequations}
The transfer matrix describes the relation between fields at $z^-$ and $z^+$ via $\vect{a}(z^+) = \vect{T}_\mathrm{trans}\vect{a}(z^-)$; we have \cite{Iakoupov2016}
\begin{equation}
    \vect{T}_\mathrm{trans} = \left(\begin{array}{cc} \vect{S}_R-\vect{S}_{RL}\vect{S}_L^{-1}\vect{S}_{LR} & \vect{S}_{RL}\vect{S}_L^{-1} \\ -\vect{S}_L^{-1}\vect{S}_{LR} & \vect{S}_L^{-1} \end{array}\right).
\end{equation}
Next, the propagation of the fields between two transducers is described by the transfer matrix
\begin{equation}\label{eq:SMS2T}
    \vect{T}_\mathrm{free} = \diag\left(e^{-(\zeta_1 - ik_1)d}, e^{-(\zeta_2 - ik_2)d}, e^{-(\zeta_1 + ik_1)d}, e^{-(\zeta_2 + ik_2)d}\right);
\end{equation}
here, $\zeta_i$ is a parameter describing losses (for simplicity, we assume equal loss in both fields, $\zeta_1 = \zeta_2$), $k_i = \omega/v_i$ is the wavenumber, and $d$ is the distance between the transducers.
Propagation through a unit cell of the array is now described by a product of the transfer matrices for the transducer and free propagation, $\vect{T}_j = \vect{T}_\mathrm{free}\vect{T}_{\mathrm{trans},j}$
and propagation through the whole array by the product of transfer matrices over all unit cells, $\vect{T} = \vect{T}_N\vect{T}_{N-1}\ldots\vect{T}_1$.
We can convert the transfer matrix $\vect{T}$ into a scattering matrix using a formula analogous to Eq.~\eqref{eq:SMS2T};
the resulting matrix characterizes transformation of arbitrary input signals by the array.

\begin{figure}
    \centering
    \includegraphics[width=0.8\linewidth]{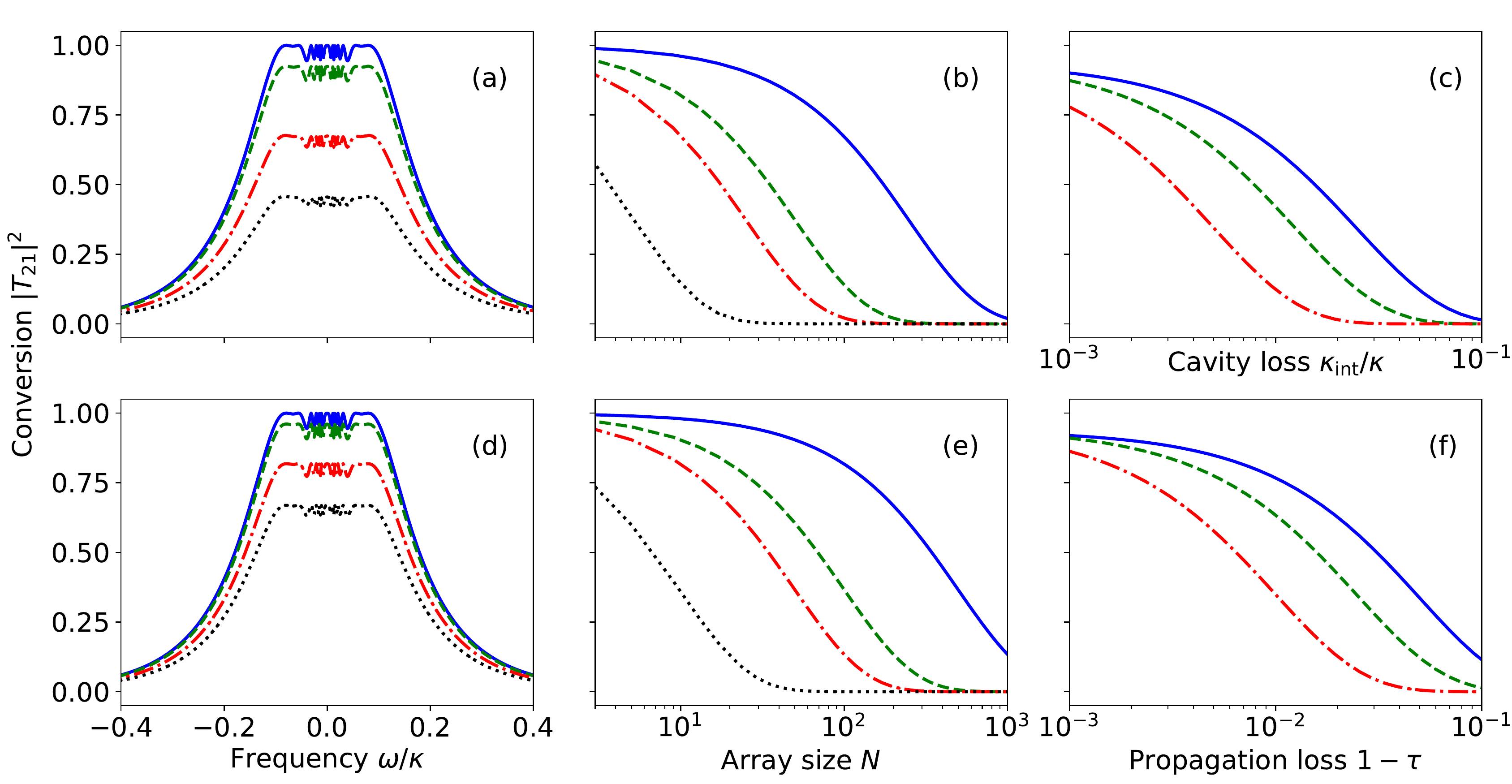}
    \caption{\label{fig:DirectLoss}
    Conversion efficiency in presence of intrinsic cavity loss (top row) and propagation loss (bottom).
        (a) Conversion spectrum with $\kappa_{\mathrm{int}} = 0$ (solid blue line), $\kappa_{\mathrm{int}} = 0.001\kappa$ (dashed green line), $\kappa_{\mathrm{int}} = 0.005\kappa$ (dot-dashed red line), and $\kappa_{\mathrm{int}} = 0.01\kappa$ (dotted black line) for an array of $N = 20$ transducers.
        (b) Conversion efficiency on resonance as a function of array size for $\kappa_{\mathrm{int}} = 0.001\kappa$ (solid blue line), $\kappa_{\mathrm{int}} = 0.005\kappa$ (dashed green line), $\kappa_{\mathrm{int}} = 0.01\kappa$ (dot-dashed red line), and $\kappa_{\mathrm{int}} = 0.05\kappa$ (dotted black line).
        (c) Conversion efficiency versus cavity loss for array size $N = 10$ (solid blue line), $N = 20$ (dashed green line), and $N = 50$ (dot-dashed red line).
        Panels (d--f) show the same for various propagation losses instead of cavity loss;
        the values are (in the same order as above) $\epsilon = 1-e^{-\zeta d} = 0,\ 0.001,\ 0.005,\ 0.01$ for panel (d)
        and $\epsilon = 0.001,\ 0.005,\ 0.01,\ 0.05$ for panel (e);
        the array sizes in panel (f) are identical with sizes used in (b).
        For all plots, the loss rates are equal for both fields, $\kappa_{1,\mathrm{int}} = \kappa_{2,\mathrm{int}} = \kappa_\mathrm{int}$, $\zeta_1 = \zeta_2 = \zeta$.}
\end{figure}

The effect of direct losses on frequency conversion is analysed in Fig.~\ref{fig:DirectLoss}.
For both intrinsic cavity loss (top row) and propagation loss (bottom), the behaviour is qualitatively the same;
cavity loss is, generally, more detrimental than propagation loss.
From the conversion spectra [panels (a,d)], we can see that losses limit the overall conversion efficiency without changing the bandwidth or spectral profile.
The following plots [conversion efficiency versus array size in panels (b,e) and versus cavity and propagation loss in panels (c,f)] show the efficiency on resonance.

Especially in the microwave field, the loss may be accompanied by thermal noise.
Since the base temperature of the device is low (in the range of 10 to 20 mK), however, the thermal occupation remains low and only slightly elevates the noise floor from the vacuum noise (by a factor of the order of unity).
In the regime of high conversion efficiency (and, by extension, low loss), the difference between vacuum and thermal baths for the microwave field remains negligible.

\begin{figure}[b!]
    \centering
    \includegraphics[width=0.75\linewidth]{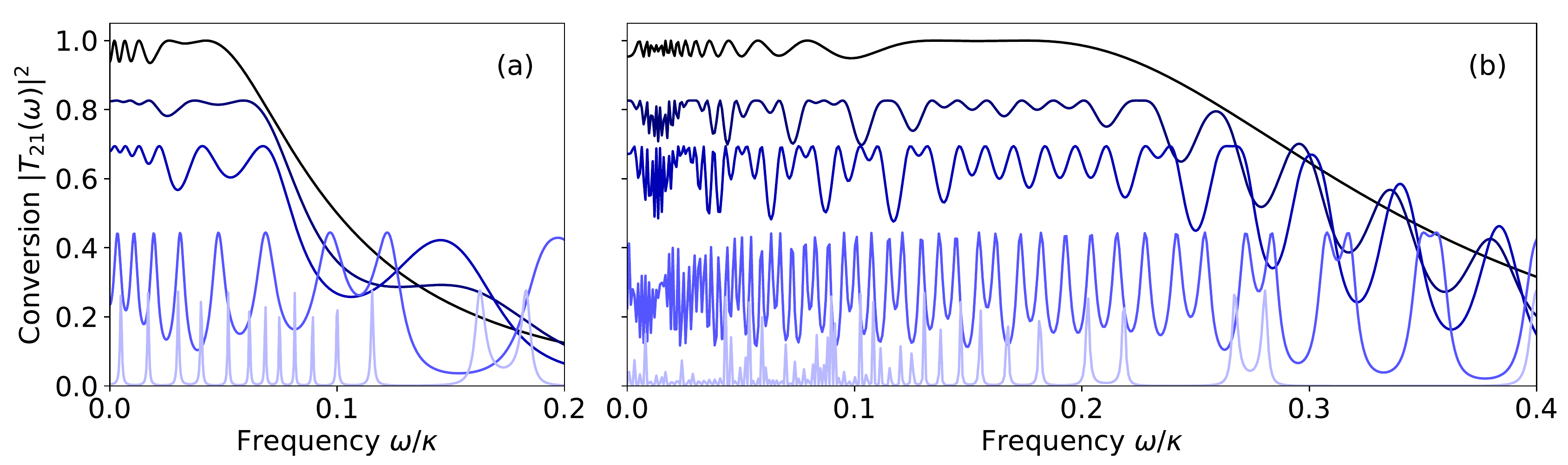}
    \caption{\label{fig:ABackScatter}
        Conversion spectra with (a) $N = 10$ and (b) $N = 50$ transducers in the presence of backscattering.
        For both panels, the backscattering rate increases from dark colours to light;
        we have $\kappa_{{L}}/\kappa_{{R}} = 0,\ 0.1,\ 0.2,\ 0.5,\ 0.9$.
    }
\end{figure}

Conversion spectra in presence of backscattering are plotted in Fig.~\ref{fig:ABackScatter} for arrays with $N = 10$ [panel (a)] and $N = 50$ transducers [panel (b)].
The backscattering rate $\kappa_{L}$ reduces the overall conversion efficiency; this decrease is independent of the array size.
Additionally, owing to the large phase shift the signal acquires during propagation through the array, the forward- and backward-propagating signals partially interfere.
This interference manifests as oscillation of the conversion efficiency with frequency.
The modulation depth depends on the backscattering rate (for equal scattering rates for backwards- and forwards-propagating fields, total destructive interference occurs) while its frequency depends on the phase shift of the signal and thus on the array size.

\end{widetext}

\end{document}